\documentclass[twocolumn,showpacs,preprintnumbers,prl,nofootinbib]{revtex4-1}
\pdfoutput=1
\usepackage{graphicx,epsfig,amsmath,amssymb,bm,slashed,wasysym,multirow}
\usepackage{color}
\usepackage{ifthen} 
\newboolean{pdflatex}
\setboolean{pdflatex}{True} 

\newboolean{articletitles}
\setboolean{articletitles}{true} 

\newboolean{uprightparticles}
\setboolean{uprightparticles}{false} 

\newboolean{inbibliography}
\setboolean{inbibliography}{false} 

\def \beq{\begin{equation}}
\def \eeq{\end{equation}}
\def \beqa{\begin{eqnarray}}
\def \eeqa{\end{eqnarray}}

\makeatletter
\DeclareRobustCommand*{\bfseries}{%
  \not@math@alphabet\bfseries\mathbf
  \fontseries\bfdefault\selectfont
  \boldmath
}
\makeatother

\usepackage{mciteplus}
\begin{document}

\preprint{} 

\title{Leptonic top-quark asymmetry predictions at LHCb}

\author{Rhorry Gauld$^a$} 

\affiliation{${}^a$\,Department of Physics, University of Oxford, 
OX1 3PN Oxford, United Kingdom}

\date{\today}

\begin{abstract}
The forward LHCb acceptance offers the possibility of measuring the top-quark pair asymmetry in a kinematic region that does not receive overwhelming dilution from the symmetric gluon-fusion channel. To investigate this possibility, two leptonic final states are identified, and analysis strategies are proposed for each channel with 14~TeV data.  Leading fixed-order predictions
are then provided for the relevant leptonic asymmetry variables in each channel. If backgrounds can be experimentally constrained, statistically, a non-zero asymmetry is estimated to be observable beyond 5$\sigma$ confidence level with the full LHCb 14~TeV data.
\end{abstract}

\maketitle

\section{Introduction} 
Top-quark production has so far not been observed at very high pseudorapidity\footnote{The LHCb collaboration has now observed top quark production~\cite{Aaij:2015mwa} with the Run-I data. This measurement is statistically limited, and the focus of this study will be for Run-II measurements where the statistical reach is extended significantly.}. 
As the LHCb detector is instrumented in this region of phase space, it is important to investigate the feasibility of measuring the properties of top-quark production with the available and expected future LHCb data.

At the LHC, top-quarks are predominantly pair produced via the strong interaction. This production channel is therefore statistically the most promising for performing precision measurements in an extreme kinematic region. Besides the statistical benefits, measurements of forwardly produced top-quark pairs are also well motivated, as they provide sensitivity to partons within the colliding protons with both moderate- and large-$x$ momentum fractions.

An interesting consequence of this kinematic sensitivity is that the relative contribution of quark-initiated subprocesses increases for forwardly produced top-quark pairs --- a result of the relative decline of the gluon with respect to the valence quark PDF at large-$x$. Within the Standard Model (SM), an asymmetry exists in the production of top-quarks pairs~\cite{Halzen:1987xd,Nason:1989zy,Kuhn:1998jr,Kuhn:1998kw} which results in an asymmetric angular distribution of top- and antitop-quarks. This asymmetry arises from quark-initiated subprocesses of the form $q\bar{q}\rightarrow t\bar{t}X$, where $X = g,\,\gamma,\,Z,\,W^{\pm}$ denotes a particular final state\footnote{By crossing symmetry, subprocesses of the form $q(g/\gamma)\rightarrow t\bar{t}q$ also contribute to the asymmetry.}. At a hadron collider, observing such a final state asymmetry relies on the presence of an asymmetry in the initial state PDFs. Asymmetry measurements at LHCb, as originally suggested in~\cite{Kagan:2011yx}, are therefore well motivated as the asymmetry between the quark and antiquark PDFs increases at large-$x$. The contribution arising from the symmetric gluon-fusion ($gg$) subprocess which dilutes the asymmetry is also relatively suppressed for forwardly produced top-quark pairs.
\footnote{This argument also applies to $c$- and $b$-quark heavy quark pair asymmetries, and was the motivation for the
LHCb measurement of $b$-quark asymmetry at 7~TeV~\cite{Aaij:2014ywa}. However, this measurement is consistent with the corresponding SM prediction~\cite{Gauld:2015qha,Murphy:2015cha}.}
%
%

Measurements of the charge asymmetry in the forward region are also experimentally well motivated, as both the CDF~\cite{Aaltonen:2012it,Aaltonen:2013vaf} and D0~\cite{Abazov:2014cca,Abazov:2014oea,Abazov:2013wxa} experiments at the Tevatron observe asymmetries larger than the corresponding predictions in the SM. Although the overall tension is rather small, there is some indication that this discrepancy increases for very forwardly (backwardly) produced (anti)top-quarks. Measurements at the LHC have already been performed~\cite{ATLAS-CONF-2012-057,Chatrchyan:2014yta,CMS-PAS-TOP-14-006,CMS-PAS-TOP-12-033}, but have been so far inconclusive in confirming or refuting the behaviour observed at the Tevatron. Although performed with higher statistics, asymmetry measurements in the central region at the LHC will remain challenging as the observable asymmetric cross section is substantially diluted by the symmetric $gg$ channel. It is worth noting that there have been several proposals~\cite{Berge:2013xsa,Aguilar-Saavedra:2014vta,Maltoni:2014zpa} which indicate that central measurements at CMS and ATLAS may avoid these overwhelming dilution effects if associated production processes or carefully selected kinematic observables are studied.

Due to the limited forward acceptance of the LHCb detector, only partial reconstruction of the full $t\bar{t}X$ final state is possible. However, as has been previously pointed out in~\cite{Kagan:2011yx}, reasonable event rates in the single-lepton final state $t\bar{t}X\rightarrow lb$ are achievable with suitable analysis cuts. An asymmetry can then be inferred by measuring the rate of $l^{+}b$ to $l^{-}b$ events. In this work, the signal and background rates for this channel are re-assessed at $\sqrt{s} = 14$~TeV where the proposed measurement becomes statistically feasible. In addition to this, measurements of the di-lepton final state $t\bar{t}X\rightarrow \mu eb$ are also discussed. Although statistically less promising, this channel opens the possibility of measuring the rapidity difference of reconstructed leptons on an event-by-event basis, and is also essentially free of background. In what follows, a set of analysis cuts are selected for each of these final states, after considering the relevant background rates, and fixed-order predictions are provided for relevant asymmetry observables at the leptonic level.

\section{Structure of prediction}
Before studying specific leptonic final states, it is important to review the structure of the asymmetry prediction. As demonstrated in~\cite{Kuhn:1998jr,Kuhn:1998kw}, the dominant contribution arises from the interference of amplitudes at next-to-leading order (NLO) which are relatively odd under interchange of final state top- and antitop-quarks --- of which the most significant contribution is the purely QCD effect of $\mathcal{O}(\alpha_s^3)$. In addition to pure QCD, mixed QCD-electroweak (-EW) effects of $\mathcal{O}(\alpha_s^2\alpha_{\rm{EW}})$, where either a single photon or $Z$ boson is present in the squared amplitude, also contribute to the asymmetry. As well as these amplitude effects, there is also a purely electroweak (EW) contribution to the asymmetry which arises at LO due to the presence of both axial and vector couplings to fermions present in $\gamma$-$Z$ and $Z$-$Z$ interference. The prediction for the asymmetry is therefore cast in terms of a Taylor series expansion in powers of the strong ($\alpha_s$) and EW ($\alpha_{\rm{EW}}$) couplings in the following way
\beq
\begin{aligned} \label{casym}
A &= \frac{\alpha_s^3 \sigma_a^{s(1)}  + \alpha_s^2 \alpha_{\rm{EW}} \sigma_a^{\rm{EW}(1)} + \alpha_{\rm{EW}}^2 \sigma_a^{\rm{EW}(0)} + \cdot \cdot \cdot}{\alpha_s^2 \sigma_s^{s(0)} + \alpha_s^3 \sigma_s^{s(1)} +\cdot \cdot \cdot} \, , \\
	&= \alpha_s \frac{\sigma_a^{s(1)}}{\sigma_s^{s(0)}} + \alpha_{\rm{EW}} \frac{\sigma_a^{\rm{EW}(1)}}{\sigma_s^{s(0)}} + \frac{\alpha_{\rm{EW}}^2}{\alpha_s^2} \frac{\sigma_a^{\rm{EW}(0)}}{\sigma_s^{s(0)}} + {\cdot \cdot \cdot} \, .
\end{aligned}
\eeq
\noindent In this notation $\sigma_{a}^{s(1)}$ and $\sigma_{a}^{\rm{EW}(1)}$ represent the purely QCD and mixed QCD-EW asymmetric effects arising at NLO, $\sigma_{a}^{\rm{EW}(0)}$ is the purely EW asymmetric effect arising at LO, and $\sigma_s^{s(0)}$ is the symmetric QCD contribution arising at LO. In the same fashion as~\cite{Kuhn:2011ri,Hollik:2011ps,Bernreuther:2012sx}, the expansion in~(\ref{casym}) is truncated to avoid terms containing $\sigma_a^{s(2)}$, which is currently unavailable for unstable top-quarks. It should be noted that this term has now been computed for stable top-quarks at the Tevatron~\cite{Czakon:2014xsa}, and is observed to be both positive and of comparable size to the aforementioned mixed QCD-EW contribution. 

As previously discussed, the considered final states never involve fully reconstructing top-quarks, and consequently the asymmetry is accessed by studying the angular distributions of leptonic top-quark decays. The decay of top-quarks is included at LO by factorising the full amplitude into production and decay amplitudes, under the assumption that the top-quarks are produced on-shell.

To obtain the $\mathcal{O}(\alpha_s^3)$ contribution to the numerator, an adapted version of the publicly available \texttt{MCFM}~\cite{Campbell:2010ff,Campbell:2012uf} program is used, which separates the individual contributions from the $u\bar{u}$, $d\bar{d}$, $ug$, and $dg$ subprocesses. The $\mathcal{O}(\alpha_s^2 \alpha_{\rm{EW}})$ can then be (approximately) obtained from these results by applying a rescaling of couplings and colour factors. From diagram inspection, the ratio of the $\mathcal{O}(\alpha_s^2\alpha_{\rm{EW}})$ to the $\mathcal{O}(\alpha_s^3)$ results for $q\bar{q}$- and $qg$-initiated states are given by
\beq \label{QED}
R^{X}_{q\bar{q}}(\mu) = \frac{36 Q^{X}_q Q^{X}_t \alpha_e}{5 \alpha_s}  \, , \quad
R^{X}_{qg}(\mu) = \frac{24 Q^{X}_q Q^{X}_t \alpha_e}{5 \alpha_s} \, ,
\eeq
\noindent where $Q^X_q$ and $Q^X_t$ refer to the relevant quark and top-quark charges and $\alpha_e$ denotes the electromagnetic coupling. The electromagnetic contribution is found by including the relevant electromagnetic quark charges $Q^{e}$. The weak contribution is found with the replacement $Q^w = (2\tau^3-4 s_w^2 Q^{e})/4 s_w c_w$, where $\tau^3$ and $(c)s_w$ are weak isospin and the (co)sine of the weak mixing angle $\theta_w$ respectively. The first replacement is exact\footnote{Excluding the negligible contribution from the $q\gamma$-initiated subprocess.}, and the second replacement is valid under the assumption $m_Z/\sqrt{\hat{s}} \ll 1$. For the forward region of phase space considered in this work, this approximation is estimated to account for $\simeq$ 0.95 of the full effect --- this is found by comparing the full and rescaled result for stable top-quarks when both top-quarks are within the LHCb acceptance. The $\mathcal{O}(\alpha_{\rm{EW}}^2)$ contribution is also available in \texttt{MCFM}, and has been extended to include hadronic $W$ boson decays, which is necessary for the single-lepton final state. The numerical accuracy of the results is estimated to be $\mathcal{O}(1\%)$, which is found by generating 100 statistically independent samples and calculating the standard deviation. In the $qg$ channels the accuracy is slightly poorer, however the contribution of this channel to the total asymmetry is relatively always below 10\%. The running of $s_w^2(m_Z) = 0.231$, and $\alpha_e(m_Z) = 1/128$ are considered at one-loop.

Finally, the dependence on the choice of PDFs and scales is evaluated in the following way. The numerator of each asymmetry is computed with NNPDF2.3 NLO PDFs with $\alpha_s (m_Z) =$ 0.119~\cite{Ball:2011mu}. The denominator is then computed with the LO 0.119, LO 0.130, and NLO 0.119 NNPDF2.3 PDFs. A scale uncertainty is evaluated by simultaneously computing the numerator and denominator of each asymmetry for a specific scale choice $\mu_F = \mu_R = \mu = \{m_t/2,m_t,2m_t\}$. The central value is then found by averaging these three predictions, and an uncertainty is associated to the total envelope. The top mass is fixed at $m_t$ = 173.25~GeV throughout.

\section{Single-lepton asymmetry}
The single-lepton final state refers to the partially reconstructed process $t\bar{t}\rightarrow lbX$, in which a single lepton and $b$-jet are registered by the detector. A differential asymmetry can then be inferred by measuring the rate of $l^+$ to $l^-$ tagged events as
\beq \label{LHCbAleprate}
\frac{d A^{l}}{d \eta_l} = \left( \frac{d \sigma^{l^+b}/ d \eta_l - d \sigma^{{l^-b}}/ d \eta_l}{d \sigma^{l^+b}/d\eta_l+d \sigma^{l^-b}/ d \eta_l} \right) \,,
\eeq
\noindent where both the $b$-jet and lepton are required to be within the LHCb kinematic acceptance of $2.0 < \eta < 4.5$. Before studying the properties of this final state, it is necessary to include analysis cuts to manage the various sources of background. 

The main backgrounds are identified as single top, $W+$($b$)jets, $Z+$($b$)jets, and QCD. In accordance with LHCb trigger requirements, a minimum $p_T$ of 20~GeV is required for all leptons. It is also necessary to introduce an isolation criterion $\Delta R (l^{\pm}\,,\mathrm{jet}) \geq R$, which ensures the QCD contamination is negligible~\cite{Kagan:2011yx}. To apply this isolation, jets are clustered with the anti-$k_t$ algorithm~\cite{Cacciari:2008gp} with the choice $R$ = 0.5. Events in which two oppositely charged leptons simultaneously pass these analysis cuts are vetoed, and are considered in the di-lepton analysis. Throughout the single-lepton analysis, a $b$-tagging mis-tag rate of 1.4\% is applied to light jets --- this is motivated by LHCb analyses which suggest a mis-tag rate of 1\% with an associated efficiency of 70\% is achievable~\cite{Barter:1610595,Aaij:2015yqa}. A $p_T$ cut of 60~GeV is placed on this $b$-jet.

\begin{figure}[t!]
\begin{center}
\makebox{\includegraphics[width=\columnwidth]{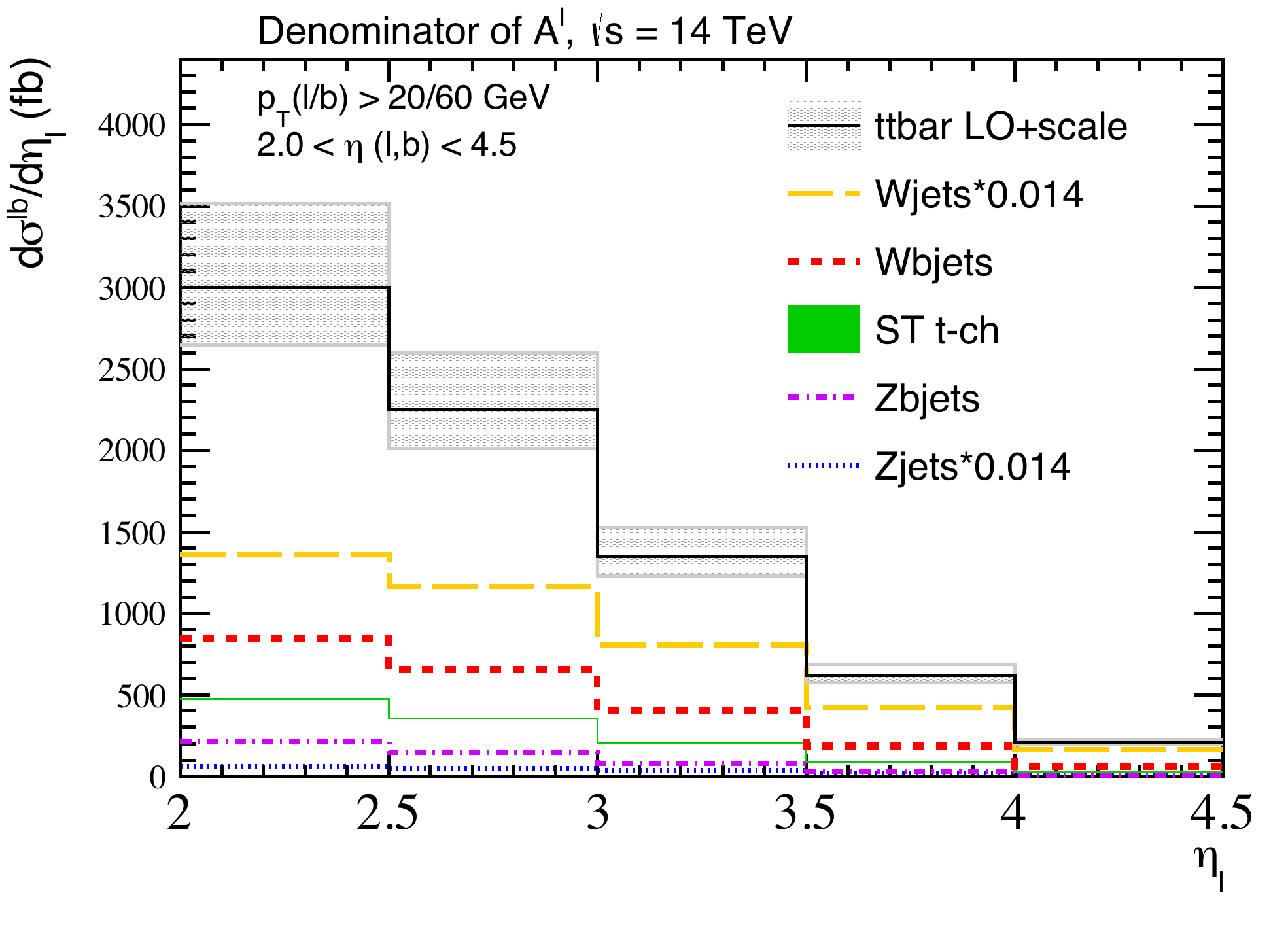}}
\end{center}
\vspace{-5mm}
\caption{Stacked contributions from signal and background processes to the symmetric cross section in the single-lepton channel at 14~TeV. Analysis cuts and relevant efficiencies have been applied to all processes. See text for details.}
\label{RateDenom}
\end{figure}

The contribution from signal and background to the symmetric cross section, including the discussed analysis cuts and efficiencies, is shown in Fig.~\ref{RateDenom}. Note that the contribution from each process has been stacked. The background samples are simulated using \texttt{POWHEG}~\cite{Alioli:2010qp,Alioli:2009je,Frederix:2012dh,Oleari:2011ey} with the central CT10w PDF set~\cite{Lai:2010vv} and then subsequently matched to \texttt{Pythia8176}~\cite{Sjostrand:2007gs} --- the only exception is $Z+b$jets where the matrix element is produced using \texttt{MadGraph5}~\cite{Alwall:2014hca} with CTEQ6ll. In the case of single top, only the $t$-channel process is considered, and an uncertainty is associated to the difference between 4- and 5-flavour scheme predictions. The 4-flavour inclusive cross section is also normalised to that of the 5-flavour prediction. For all (N)LO+PS background samples, jet reconstruction is performed with the \texttt{FastJet} software~\cite{Cacciari:2011ma}, and $b$-jets are found by matching $b$-quarks to jets at the parton level. The shown $t\bar{t}$ sample in Fig.~\ref{RateDenom} is generated at LO with NLO 0.119 NNPDF2.3 PDFs. In this work, the signal process is studied with NNPDF PDFs as they provide updated sets at (N)LO with varying choices of $\alpha_s(m_Z)$. This is important for evaluating the uncertainty of the signal asymmetry prediction, which is considered by computing the denominator with these differing PDFs. The background samples are taken from previous work~\cite{Gauld:2013aja}.

The contributions to the inclusive asymmetry, with the discussed analysis cuts applied, from the various $t\bar{t}$ subprocesses are now provided. The prediction for the numerator at various scale choices is provided in Table~\ref{t:Nrate14TeV}, while the corresponding denominator and asymmetry predictions are provided in Table~\ref{t:Arate14TeV}.

\renewcommand*{\arraystretch}{1.0}
\begin{table}[!h]
  \begin{tabular}{@{} c|c|c|c|c@{}}
    \multicolumn{2}{c|}{$N^{l}$ (fb)}	& $\mu = m_t/2$	& $\mu = m_t$		& $\mu = 2 m_t$ \\
    \hline \hline
    	\multirow{4}{*}{$\mathcal{O}(\alpha_s^3)$}	& $u\bar{u}$			& 55.62		& 40.84		& 31.56		\\
	 									& $d\bar{d}$			& 23.15		& 16.99		& 13.05	\\
						 				& $ug$				& 1.79		& 1.02		& 0.65		\\
						 				& $dg$				& 0.72		& 0.45		& 0.26		\\ \hline
	\multicolumn{2}{r|}{$\approx\mathcal{O}(\alpha_s^2\alpha_{\rm{EW}})$}			& 9.72		& 7.90		& 6.67		\\ \hline	
	\multicolumn{2}{r|}{$\mathcal{O}(\alpha_{\rm{EW}}^2)$}			& 0.81		& 0.78		& 0.77		\\ \hline
	\multicolumn{2}{c|}{Total}								& 91.80		& 67.96		& 52.95 		\\ \hline
  \end{tabular}
  \caption{Signal contribution the numerator of the inclusive leptonic rate asymmetry at 14~TeV. The analysis cuts discussed in the text have been applied.}
  \label{t:Nrate14TeV}
\end{table}

\renewcommand*{\arraystretch}{1.0}
\begin{table}[!h]
\centering
  \begin{tabular}{@{} c|c|c|c|c@{}}
     			&  \multicolumn{3}{c|}{$D^{l}$ (fb), 14~TeV}	&  \\ \hline
    PDF		& $\mu = m_t/2$		& $\mu = m_t$		& $\mu = 2 m_t$ & $A^{l}$ (\%) \\
    \hline \hline
    NLO 119	& 	4626		& 3512		& 2742	& 1.95 (3)	 \\  
    \,\,\,\,LO	 119	& 	6225		& 4663		& 3586	& 1.47 (1)	\\
    \,\,\,\,LO	 130 	& 	6761		& 4961		& 3752	& 1.38 (3) 	\\
      \end{tabular}
  \caption{Signal contributions to the denominator and leptonic rate asymmetry at 14~TeV. The analysis cuts and efficiencies discussed in the text have been applied.}
  \label{t:Arate14TeV}
\end{table}

The resultant scale uncertainty, evaluated from simultaneous scale variation of both numerator and denominator, is extremely small in all cases. The $\mathcal{O}(\alpha_s^3)$ contribution to the numerator exhibits similar scale dependence to the denominator of the asymmetry  (LO symmetric QCD) and introduces an artificial cancellation of the evaluation of possible higher-order effects. On the other hand, the choice of PDFs used in the evaluation of the LO symmetric QCD cross section has a large impact on the asymmetry. This is mainly a consequence of the difference in the behaviour of the gluon PDF for values of $x >$ 0.1 for LO and NLO PDFs. In this region, the considered NLO gluon PDF is substantially softer in comparison to both LO 119 and LO 130 gluon PDFs. When computing the symmetric NLO cross section, higher-order corrections and the presence of the $qg$-initiated subprocess compensate this effect. The absence of these effects in the computation of the LO cross section results in an underestimation of the symmetric cross section $\sigma_s^s(0)$ --- hence an overestimation of the asymmetry. As a conservative approach, the resultant asymmetry in all three cases is provided. Recent work~\cite{Czakon:2014xsa,Kidonakis:2015ona} has shown that this uncertainty is reduced with the inclusion of higher-order terms.

\begin{figure}[t!]
\begin{center}
\makebox{\includegraphics[width=\columnwidth]{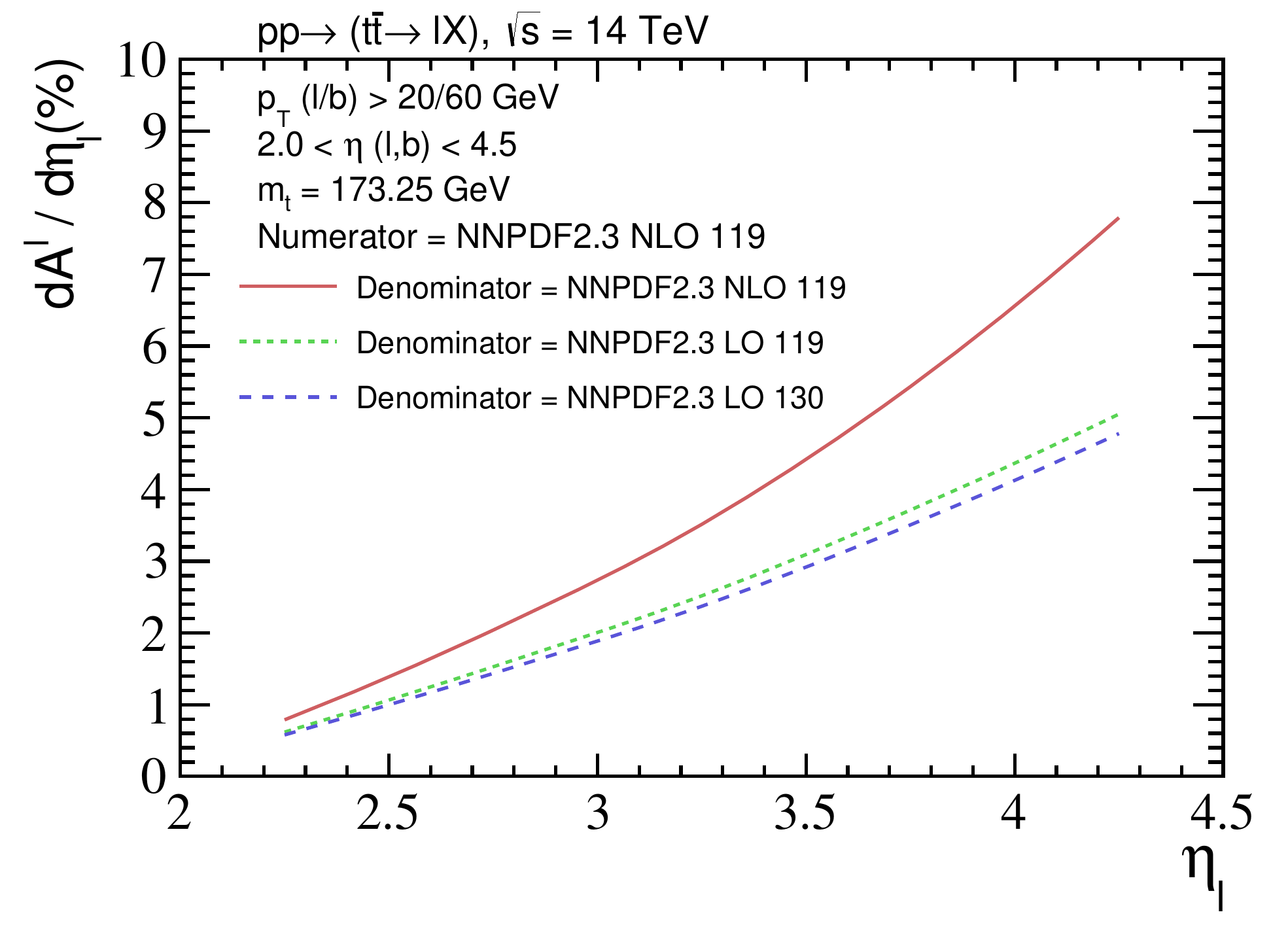}}
\end{center}
\vspace{-5mm}
\caption{Differential leptonic rate asymmetry as a function of lepton pseudorapidity at 14~TeV. The choice of analysis cuts, and PDFs used for the computation of the numerator and denominator are highlighted.}
\label{DiffLepRate}
\end{figure}

The central differential leptonic rate asymmetry is presented as function of lepton pseudorapidity in Fig.~\ref{DiffLepRate}. The dependence of the resultant asymmetry on the choice of PDFs used for the computation of the denominator has also been highlighted. The uncertainty due to scale variation is negligible and therefore not included.
%

\begin{figure}[ht!]
\begin{center}
\makebox{\includegraphics[width=\columnwidth]{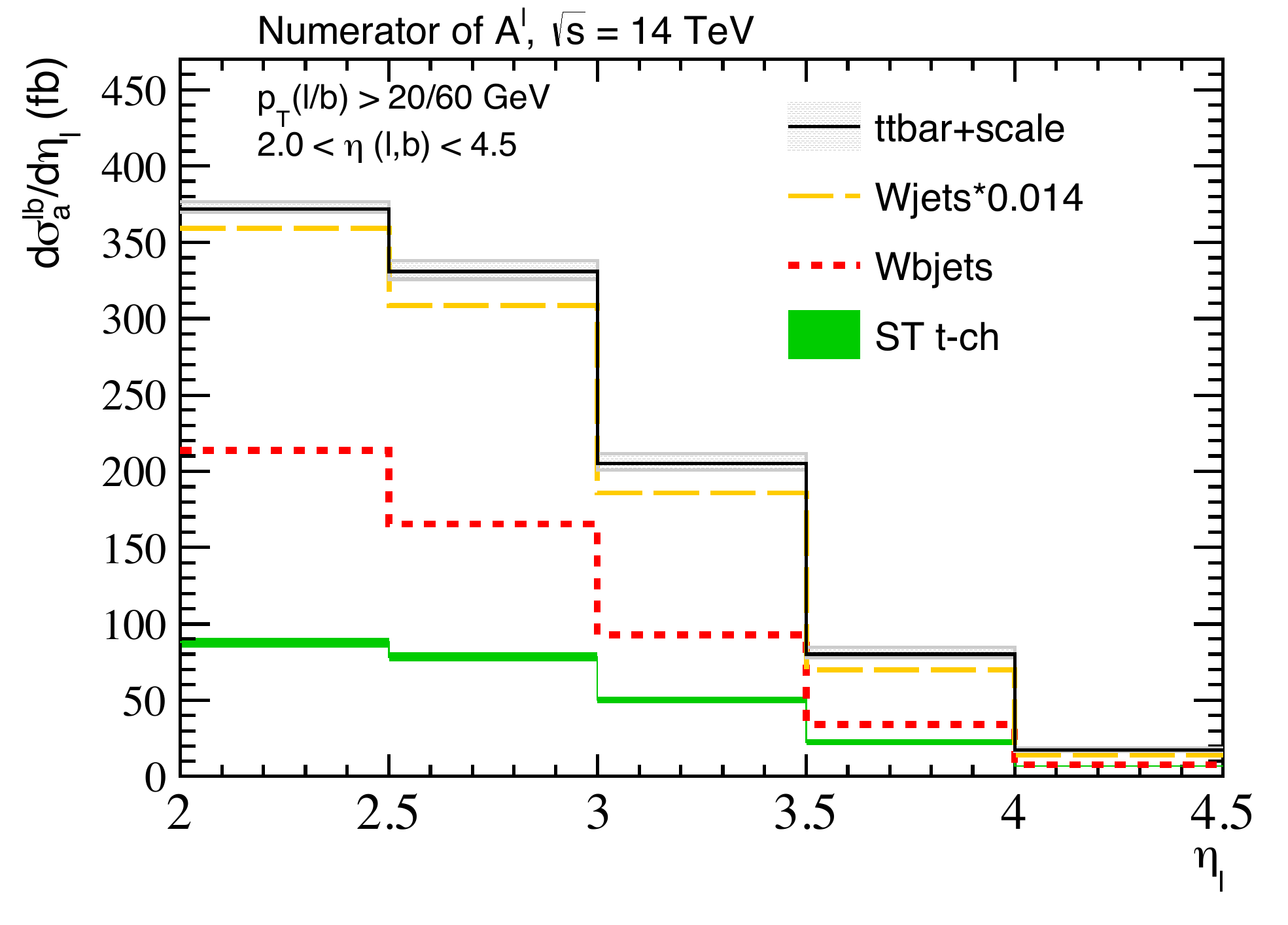}}
\end{center}
\vspace{-5mm}
\caption{Stacked signal and background contributions to the numerator of the total leptonic rate asymmetry at 14~TeV.}
\label{numlj}
\end{figure}

The signal contribution to the asymmetry is significant, particularly at large $\eta_l$ where the asymmetry reaches (3-8)\%. To experimentally extract this signal, it is however necessary to precisely know the background contribution to the asymmetry. This is demonstrated in Fig.~\ref{numlj}, where the contributions from both signal and background processes to the numerator of the asymmetry are shown.

The background contributions depend on the $b$-tagging mis-tag rate and associated efficiency in a non-trivial way, which will ultimately limit the precision of an asymmetry measurement in this channel. In principle, an experimental background fit performed across several different final states --- such as $lj$, $lbj$, $lbb$ --- should provide adequate knowledge of the backgrounds for the considered $lb$ final state. Assuming these experimental uncertainties are under control, the statistical feasibility of such a measurement can be estimated by considering the background-subtracted $t\bar{t}$ sample. This is computed according to $\delta A = \sqrt{(1-A^2)/N}$, where the number of events $N$ is found by applying a lepton efficiency of 75\% (an approximate trigger effect), a $b$-tagging efficiency of 70\%, and assuming an integrated luminosity of 50~fb$^{-1}$ --- the expected data collected by 2030~\cite{Bediaga:1443882}. The statistical significance is large-$x$computed for the three scenarios considered in Table~\ref{t:Arate14TeV} by computing the expected number of events for the scale choice $\mu = m_t$. Even in the worst case scenario (the lower cross section), the expected number of signal events is $\sim 1e5$. In all three cases, the statistical significance ($A/\delta A$) of measuring a non-zero asymmetry exceeds $\sim5\sigma$. It should be noted that this analysis is rather naive, in that perfect background subtraction has been assumed. However, it indicates that such an asymmetry measurement is statistically possible with the expected data if backgrounds can be controlled.

\section{Di-lepton asymmetry}
With the large amount of data expected by 2030, measurements in the di-lepton channel at LHCb also become feasible. As previously mentioned, by considering the partially reconstructed final state $t\bar{t} \rightarrow \mu ebX$, it becomes possible to measure the rapidity difference of reconstructed leptons on an event-by-event basis. A differential asymmetry can then be inferred by measuring a forward-backward asymmetry as
\beq \label{LHCbAlepfb}
\frac{d A^{ll}_{fb} }{d \Delta_{y} } = \frac{ \left(  d \sigma^{\mu eb} (\Delta_{y} > 0)- d \sigma^{\mu eb} (\Delta_{y} < 0) \right)/d \Delta_{y}}{d \sigma^{\mu eb}/ d \Delta_{y}}  \, ,
\eeq
\noindent where $\Delta_{y} = y_{l^+}-y_{l^-}$, and both leptons and the $b$-jet are within the acceptance $2.0 < \eta < 4.5$. The choice of opposite flavour leptons is required to remove, the otherwise overwhelming, $Z$ background processes. With this requirement in place, the main backgrounds are identified as $Z\rightarrow \tau\tau$, $WW$, $WZ$, $tW$, and QCD. In a similar fashion to the single-lepton analysis, leptons are required to be isolated, and to have a minimum $p_T$ of 20~GeV. In this analysis, a $p_T$ cut of 20~GeV is also placed on the $b$-jet, and a looser $b$-tagging mis-tag rate of 5\% is assumed in favour of efficiency.

\begin{figure}[t!]
\begin{center}
\makebox{\includegraphics[width=\columnwidth]{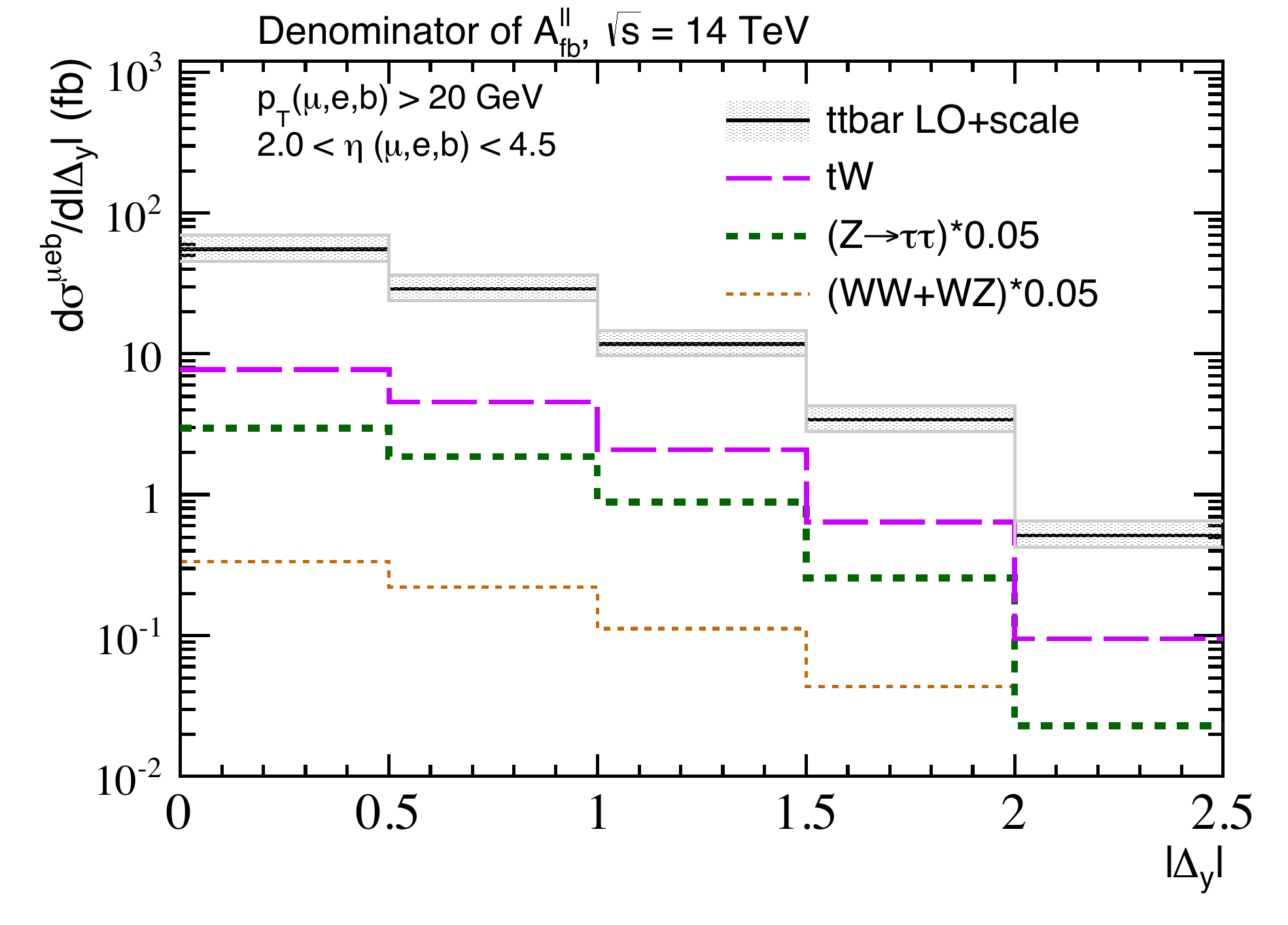}}
\end{center}
\vspace{-5mm}
\caption{Stacked contributions from signal and background processes to the symmetric cross section in the di-lepton channel at 14~TeV. Analysis cuts and relevant efficiencies have been applied to all processes. See text for details.}
\label{AfbDenom}
\end{figure}

The contribution from signal and background to the symmetric cross section is shown in Fig.~\ref{AfbDenom}. The background samples are simulated using \texttt{POWHEG}~\cite{Melia:2011tj,Re:2010bp} with the central CT10w PDF set, and then subsequently matched to \texttt{Pythia8176}. The QCD background, which is expected to arise from multi-jet production, is not considered in this study. It is possible to account for this background experimentally by measuring the event rate and kinematic distributions of same-sign $\mu$ and $e$ leptons. Internal studies with the 8~TeV data at LHCb indicate that, after isolation and impact parameter cuts, the QCD contribution is expected to be below 10\% of the $t\bar{t}$ signal~\cite{Brown:1635683}.

Following the procedure adopted for the single-lepton final state, the prediction for the numerator at various scale choices is provided in Table~\ref{t:Nafblep14TeV}, while the corresponding denominator and asymmetry predictions are provided in Table~\ref{t:Afblep14TeV}.

\renewcommand*{\arraystretch}{1.0}
\begin{table}[!h]
  \begin{tabular}{@{} r|c|c|c|c@{}}
    \multicolumn{2}{c|}{$N^{ll}_{fb}$ (fb)}	& $\mu = m_t/2$	& $\mu = m_t$		& $\mu = 2 m_t$ \\
    \hline \hline
    	\multirow{4}{*}{$\mathcal{O}(\alpha_s^3)$}	& $u\bar{u}$		& 0.977			& 0.709		& 0.536		\\
	 									& $d\bar{d}$		& 0.344			& 0.239		& 0.181		\\
						 				& $ug$			& 0.095			& 0.070		& 0.045		\\
						 				& $dg$			& 0.031			& 0.021		& 0.013		\\ \hline
	\multicolumn{2}{r|}{$\approx \mathcal{O}(\alpha_s^2\alpha_{\rm{EW}})$}& 0.187			& 0.153		& 0.126		\\ \hline
	\multicolumn{2}{r|}{$\mathcal{O}(\alpha_{\rm{EW}}^2)$}			& 0.006			& 0.005		& 0.005		\\ \hline	
	\multicolumn{2}{c|}{Total}								& 1.642			& 1.198		& 0.907	\\ \hline
  \end{tabular}
  \caption{Signal contribution to the numerator of the inclusive leptonic forward-backward asymmetry at 14~TeV. The analysis cuts discussed in the text have been applied.}
  \label{t:Nafblep14TeV}
\end{table}

\renewcommand*{\arraystretch}{1.0}
\begin{table}[ht!]
\centering
  \begin{tabular}{@{} c|c|c|c|c@{}}
     			&  \multicolumn{3}{c|}{$D^{ll}_{fb}$ (fb), 14~TeV}	&  \\ \hline
    PDF		& $\mu = m_t/2$		& $\mu = m_t$		& $\mu = 2 m_t$ & $A^{ll}_{fb}$ (\%) \\
    \hline \hline
    NLO 119	& 110.4		& 85.0		& 67.4	& 1.41 (8)	 \\    
    \,\,\,\,LO	 119	& 160.7		& 120.7		& 93.3	& 0.99 (3)	\\
    \,\,\,\,LO	 130 	& 176.6		& 130.0		& 98.8	& 0.92 (1) 	\\
  \end{tabular}
  \caption{Signal contribution to the denominator and leptonic forward-backward asymmetry at 14~TeV. The analysis cuts and efficiencies discussed in the text have been applied.}
  \label{t:Afblep14TeV}
\end{table}

%
%
The statistical significance of a measurement in this channel is also estimated applying the procedure adopted in the single-lepton analysis. Under the same assumptions, with the exception of a looser $b$-tagging efficiency of 90\%, leads to $\delta A_{\mathrm{stat.}} \approx$ (1.5-1.9)\%, which slightly exceeds the corresponding prediction of $A^{ll}_{fb} =$ (0.9-1.4)\%. Although statistically limited, this final state is still interesting as it is extremely clean. Cross section measurements in this final state will also become important for constraining the gluon PDF at large-$x$. For the suggested analysis cuts, the average incoming parton momentum fractions are $\langle x_1 \rangle =$ 0.37 and $\langle x_2 \rangle =$ 0.01. Within a year of running at 14~TeV, the expected number of signal events is $\sim$ 300 corresponding to a statistical uncertainty of 6\%. As discussed in~\cite{Gauld:2013aja}, such experimental precision is already sufficient to place constraints on the gluon PDF beyond $x = 0.1$.

\section{Conclusions}
Leptonic top-quark asymmetry measurements at LHCb with the full data at 14~TeV are statistically feasible. A Measurement in the single-lepton final state is particularly promising, but will require careful experimental consideration of backgrounds --- in particular, precision cross section and asymmetry measurements of the $W$+jets process within the LHCb acceptance~\cite{Farry:2015xha} will be crucial.
Within the first year of data taking at 14~TeV, there will be sufficient statistics to warrant a dedicated asymmetry measurement. In the considered di-lepton final state, background rates are extremely low which allows for clean differential cross section measurements to be performed. Asymmetry measurements in this channel should also be pursued, however it is likely they will be statistically dominated even with the full data.
 
\section{Acknowledgements}
I am grateful to Stehphen Farry, Ulrich Haisch, Jernej Kamenik, and Emanuele Re for providing useful comments. This work was completed whilst receiving the support of an STFC Postgraduate Studentship.

\section{Appendix}
The contribution of the various sources of background to the numerator and the denominator of the single-lepton ($lb$) asymmetry are provided in Table~\ref{denom} and Table~\ref{numer} respectively. The central value of the signal process generated with NLO 119 PDFs, which is shown in Fig.~\ref{RateDenom} and Fig.~\ref{numlj}, is also included. In Table~\ref{dilep}, the central value of signal (generated with NLO 119) and the $tW$ background contribution to the denominator of the dilepton ($\mu eb$) asymmetry are provided --- the other backgrounds are negligible.

\renewcommand*{\arraystretch}{1.0}
\begin{table}[ht!]
\centering
  \begin{tabular}{@{} c|c|c|c|c|c@{}}
     $D^l$ (fb)			&  \multicolumn{5}{c}{Lepton pseudorapidity bins}	 \\ \hline
    subprocess			& 2.0-2.5		& 2.5-3.0		& 3.0-3.5		& 3.5-4.0		& 4.0-4.5 \\ \hline
    $t\bar{t}(\mu=m_t)	$	& 1638.5 		& 1086.7		& 544.8		& 195.2  		& 46.3  \\
    $W$jets($\times$0.14)	& 518.9		& 507.3 		& 400.8 		& 240.6 		& 103.6  \\    
    $Wb$jets			& 366.4		& 300.5  		& 200.3 		& 100.3  		& 36.1 \\ 
    ST, t-ch				& 264.0 		& 209.3 		& 124.5 		& 52.5 		& 15.0  \\     
    $Zb$jets			&153.1 		& 98.0		& 44.9		& 15.4		& 2.9 \\    
    $Z$jets($\times$0.14)	& 58.8		& 48.6 		& 34.6 		& 17.4 		& 6.5  \\    
    \hline \hline
  \end{tabular}
  \caption{Denominator of the asymmetry $A^l$ for the $lb$ final state. Contributions from signal and background are provided. Note that a mis-tag rate of 0.14 has been subprocesses containing light jets.}
  \label{denom}
\end{table}

\renewcommand*{\arraystretch}{1.0}
\begin{table}[ht!]
\centering
  \begin{tabular}{@{} c|c|c|c|c|c@{}}
     $N^l$ (fb)			&  \multicolumn{5}{c}{Lepton pseudorapidity bins}	 \\ \hline
    subprocess			& 2.0-2.5		& 2.5-3.0		& 3.0-3.5		& 3.5-4.0		& 4.0-4.5 \\ \hline
    $t\bar{t}(\mu=m_t)	$	& 12.5 		& 22.2		& 19.2		& 10.5 		& 3.6 \\
    $W$jets($\times$0.14)	& 145.6		& 143.1		& 93.0		& 35.9		& 6.4 \\    
    $Wb$jets			& 126.0		& 87.2 		& 42.6		& 11.5 		& 0.4 \\ 
    ST, t-ch				& 87.8 		& 78.3 		& 50.3 		& 22.5 		& 6.9 \\   
    \hline \hline
  \end{tabular}
  \caption{Numerator of the asymmetry $A^l$ for the $lb$ final state. Contributions from both signal and background are provided. Note that a mis-tag rate of 0.14 has been subprocesses containing light jets.}
  \label{numer}
\end{table}

\renewcommand*{\arraystretch}{1.0}
\begin{table}[ht!]
\centering
  \begin{tabular}{@{} c|c|c|c|c|c@{}}
     $D^{ll}_{fb}$ (fb)			&  \multicolumn{5}{c}{Lepton rapidity difference bins}	 \\ \hline
    subprocess				& 0.0-0.5		& 0.5-1.0		& 1.0-1.5		& 1.5-2.0		& 2.0-2.5 \\ \hline
    $t\bar{t}(\mu=m_t)	$		& 47.7		& 24.4		& 9.7			& 2.8			& 0.4  \\
    $tW$					& 4.7			& 2.7			& 1.2	 		& 0.4			& 0.1  \\    
    \hline
  \end{tabular}
  \caption{Denominator of the asymmetry $A^{ll}_{fb}$ for the $\mu eb$ final state. Contributions from both signal and background are provided.}
  \label{dilep}
\end{table}

\bibliographystyle{BIB}
\bibliography{shortyPRL}

\end{document}